\documentclass{article}  
\usepackage{graphicx}
\usepackage[varg]{txfonts}

\begin{document}
   \title{Detection of H$_2$D$^+$ in a massive prestellar core in Orion B
 \thanks{This publication is based on data acquired with the 
          Atacama Pathfinder Experiment (APEX). APEX is a collaboration
          between the Max-Planck-Institut f\"ur Radioastronomie, the
          European Southern Observatory, and the Onsala Space
          Observatory.}}

   \author{J. Harju$^1$ \and  L.K. Haikala$^1$ \and  
           K. Lehtinen$^1$ \and M. Juvela$^1$ \and
           K. Mattila$^1$ \and O. Miettinen$^1$ \and
           M. Dumke$^2$ \and R. G\"usten$^3$ \and  
           L.-\AA. Nyman$^2$}
         
\maketitle

              {$^1$Observatory, P.O. Box 14, 
              FIN-00014 University of Helsinki, Finland    
              
               {$^2$European Southern Observatory, Alonso de Cordova 3107, 
                Santiago, Chile}
               
               {$^2$Max-Planck-Institut f\"ur Radioastronomie, Auf dem
               H\"ugel 69, 53121 Bonn, Germany} }

% \abstract{}{}{}{}{} 
% 5 {} token are mandatory
 
  \begin{abstract} 

  {\sl Aims.}
  The purpose of this study is to examine the prediction that the
  deuterated H$_3^+$ ion, H$_2$D$^+$, can be found exclusively in the
  coldest regions of molecular cloud cores.  This is also a
  feasibility study for the detection of the ground-state line of {\sl
  ortho}-H$_2$D$^+$ at 372 GHz with APEX.

  {\sl Methods.}
  The $(1_{10} \rightarrow 1_{11})$ transition of H$_2$D$^+$ at 372
  GHz was searched towards selected positions in the massive star
  forming cloud OriB9, in the dark cloud L183, and in the low- to
  intermediate mass star-forming cloud R CrA.

  {\sl Results.}
  The line was detected in cold, prestellar cores in the regions of
  OriB9 and L183, but only upper limits were obtained towards other
  locations which either have elevated temperatures or contain a newly
  born star. The H$_2$D$^+$ detection towards OriB9 is the first one
  in a massive star-forming region. The fractional {\sl
  ortho}-H$_2$D$^+$ abundances (relative to H$_2$) are estimated to be
  $\sim 1\,10^{-10}$ in two cold cores in OriB9, and $3\,10^{-10}$ in
  the cold core of L183.

  {\sl Conclusions.}
  The H$_2$D$^+$ detection in OriB9 shows that also massive star
  forming regions contain very cold prestellar cores which probably
  have reached matured chemical composition characterized, e.g., by a
  high degree of deuterium fractionation.  Besides as a tracer of the
  interior parts of prestellar cores, H$_2$D$^+$ may therefore be used
  to put contraints on the timescales related to massive star
  formation.

\end{abstract}

%
%________________________________________________________________

\section{Introduction}

The trihydrogen ion, H$_3^+$, is supposed to become the principal
carrier of positive charge in the centres of cold, dense cores when
'heavy' elements like C, O and N, are nearly completely depleted
(\cite{walmsley2004}). Because deuterium fractionation reactions are
favoured in cold gas, relatively large
abundances of the isotopoloques H$_2$D$^+$ and D$_2$H$^+$ are to be
expected in these objects. This has been also confirmed by
observations (\cite{caselli2003}; \cite{vastel2004}).
While infrared absorption spectroscopy of H$_3^+$ can be used 
to extract vital information on the gas columns towards
infrared sources (e.g. \cite{mccall1999}), the rotational lines of
H$_2$D$^+$ and D$_2$H$^+$ probe the physical conditions of cold,
prestellar cores. The abundance of H$_3^+$ and its deuterated forms
depend on the cosmic ray ionization rate of H$_2$, and on the
abundances of destructing agents: electrons, gaseous neutral species
like CO and N$_2$, and negatively charged grains
(e.g. \cite{caselli2003}; \cite{walmsley2004}).
Furhermore, the H$_3^+$ abundance and the H$_2$D$^+$/H$_3^+$ abundance ratio
depend heavily on the {\sl ortho:para} ratio of H$_2$, which in turn
is a function of time and density (\cite{pineaudesforets1991}; 
\cite{flower2006}).

The ground-state $(1_{10} \rightarrow 1_{11})$ transition of {\sl
ortho}-H$_2$D$^+$ lies between adjacent atmospheric O$_2$ and H$_2$O
absorption lines and its observation requires extremely good
conditions.  A reasonable limit is that the precipitable water vapour
content of the atmosphere, PWV, is less than 0.5 mm, which can be
achieved at high-altitude observatories only. H$_2$D$^+$ has been
previously detected from Mauna Kea towards a protostellar core
(\cite{stark1999}), and in a small number of prestellar dark cloud
cores (\cite{caselli2003}; \cite{stark2004}; \cite{vastel2006}).

In this {\em Letter} we report on the first H$_2$D$^+$ observations
with the Atacama Pathfinder Experiment, APEX, during its Science
Verification periods in July, October and November 2005. The main goal
of this series of observations was to test the feasibility of the 372
GHz line observations with this instrument. In the course of these
measurements H$_2$D$^+$ was detected in a core belonging to a
high-mass star forming region. This may open new vistas to the
chemical evolution preceding the collapse of massive stars.

\section{Source selection}

The selection of targets contains five starless cores of molecular
clouds and one massive, cold core which encloses a low-luminosity
far-infrared source.

The massive, cold core {\bf OriB9} embedding IRAS
05405-0117 in the region of Orion B is described in
\cite{caselli1994}, \cite{caselli1995}, and \cite{harju1993}. The
average ammonia linewidth is only $0.29$ kms$^{-1}$. The core does not
stand out on the $^{13}$CO and C$^{18}$O maps of \cite{caselli1995},
probably because of CO depletion. In this survey we have included the
three N$_2$H$^+(1-0)$ peaks found by \cite{caselli1994}. These are
likely to pinpoint separate clumps within the core.  The clump
associated with the IRAS source is likely to represent an early stage
of collapse at which newly born stars have not yet disturbed their
surroundings. The subsidiary clumps (OriB9 E and N) may be in
a still earlier, pre-collapse phase. Depending on the time spent in
the pre-collapse phase, chemical evolution may have resulted in a high
degree of depletion and an increased H$_2$D$^+$ abundance.

The two positions observed towards the nearby, starless dark cloud
{\bf L183} (L134N) (see e.g. \cite{pagani2005} and
references therein) correspond to the $450\mu$m continuum peaks in the
SCUBA map of \cite{kirk2005}. The southern maximum (L183-S,
the $850\mu$m emission peak) can be assigned to a very cold, dense
core with a high degree of molecular depletion. The two sources may
represent different evolutionary stages of prestellar cores
(\cite{lehtinen2003}).  Therefore, it is interesting to compare their
chemical and dynamical properties. After perfoming these observations
we learned that the H$_2$D$^+$ has been detected at several positions
along the dense ridge of L183 with the CSO by
\cite{vastel2006}.

The starless, dense clump {\bf R CrA NW} in the northwestern part of
the R Coronae Australis cloud lies close to the 1.2 mm continuum
source ``MMS10'' (\cite{chini2003}).  Our position corresponds to an
ammonia peak (Harju et al., in preparation). In this clump the kinetic
temperature is higher than in the rest of our sample, probably 
implying that depletion is less marked.

\begin{table}[ht]
\begin{center}
      \caption[]{Target positions.}
         \label{table:sourcelist}
\begin{tabular}{lllll}\hline\hline
Core      &   $\alpha_{2000}$ &  $\delta_{2000}$  &   $T_{\rm kin}$ \\
          & ($^{\rm h \, m \, s}$)&
         ($^\circ \, ^\prime \, ^{\prime\prime}$) & (K) \\ \hline
IRAS 05405-0117 & 05 43 02.5  & -01 16 23   &  10   \\
OriB9 E         & 05 43 05.2  & -01 16 23   &  10   \\
OriB9 N         & 05 43 07.8  & -01 15 03   &  10    \\
L183-N          & 15 54 08.8  & -02 51 00   &  10$^*$  \\
L183-S          & 15 54 08.8  & -02 52 38   &  7$^*$    \\
RCrA NW         & 19 01 47.7  & -36 55 15   & 15-18 \\ \hline
\end{tabular}

$^* \; T_{\rm dust}$ derived from the 450/850$\mu$m ratio
(Kirk \& Ward-Thompson 2006, private communication)
\end{center}
\end{table}

\section{Observations and data reduction}

The observations were made with APEX in July 2005 (RCrA and 
L183) and
in October and November 2005 (OriB9).  The 372421.364 MHz line of
H$_2$D$^+$ was observed in the upper side band with the APEX-2A SIS
DSB receiver. The HPBW of the telescope is about $17^{\prime\prime}$ at this
frequency.  The observing mode was position switching with the off
position $-30^\prime$ away in R.A.
The integration time for each scan was 20 seconds. A calibration
measurement was done every 10 minutes.  The first observations (RCrA)
were made with the ASC 2048 channel autocorrelator using a bandwith of
128 MHz. Because the performance of the system was unsatisfactory with
the ASC, the MPIfR Fourier transform spectrometer, FFTS, was used for
all subsequent observations. The 1 GHz bandwith of the FFTS was
divided into 16384 channels resulting in a channel width of 61 kHz
which corresponds to $\sim$ 49 ms$^{-1}$ at the observed frequency.
The observing conditions ranged from excellent (PVW 0.2 mm, zenith
opacity 0.24 at 372 GHz) to reasonable (PVW 0.7 mm, zenith opacity 0.6
at 372 GHz). Depending on the weather and the elevation of the source,
the DSB system temperature was between 130 and 300~K.

Most of the observed spectra have ripple due to variations of sky
emission, reflections in the telescope optics and instability of the
receiver. In the data reduction the possible low frequency ripple was
first fit with a sinusoidal baseline whereafter possible higher
frequency ripple was removed by masking the corresponding frequency in
the Fourier transform. Finally, a first order baseline was
subtracted around the source velocity. 
The mirror sideband of the receiver, centred at about 360 GHz, lies
at a more transparent frequency than the signal sideband.  The
difference in the atmospheric opacity between the side bands was
estimated using an atmospheric model and was taken into
account in the calibration at the telescope. The telescope time spent
on this project is 19.5 hours.

\section{Results}

\begin{figure*}
\begin{center}
\includegraphics[angle=-90,width=16.3cm]{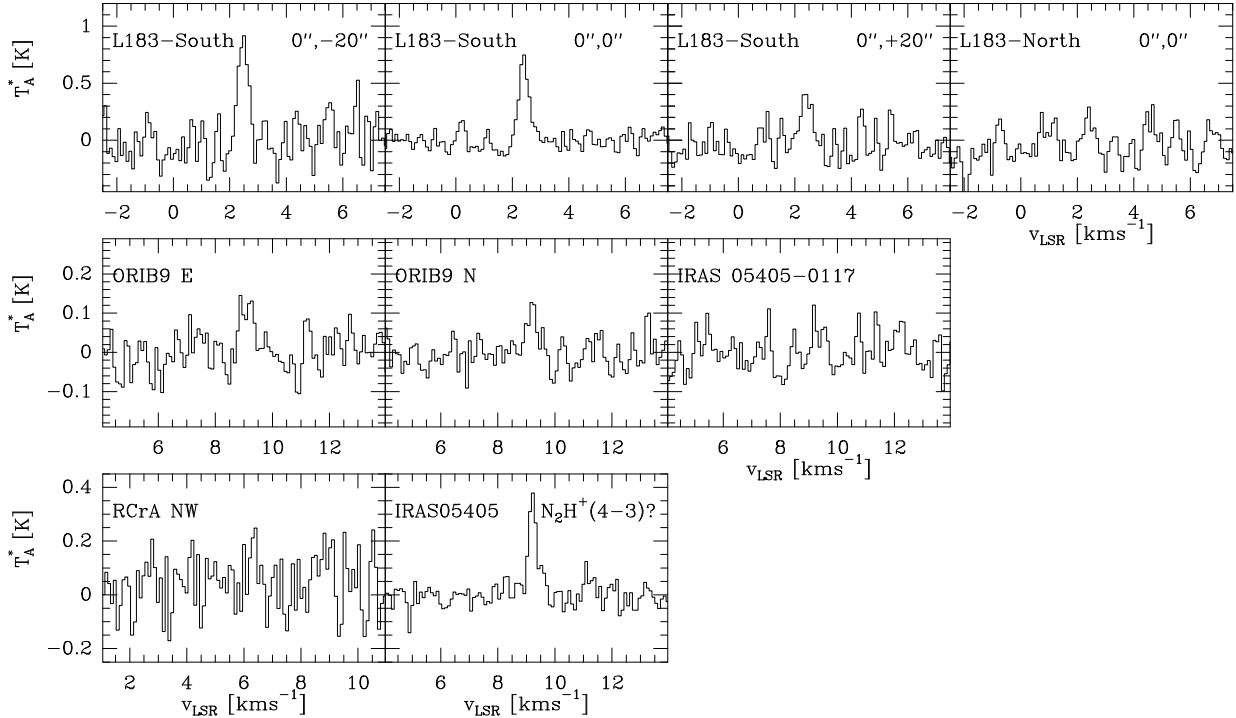}
\caption{The observed H$_2$D$^+(1_{10}-1_{11})$ spectra after
Hanning smoothing ($\Delta v = 98$ ms$^{-1}$).  {\bf Top:}
L183-S and L183-N. {\bf Middle:} OriB9-E,
OriB9-N and IRAS05405-0117.  {\bf Bottom left:} R CrA NW (ASC
spectrometer).  {\bf Bottom right:} The probable detection of
N$_2$H$^+(4-3)$ at 372.67 GHz towards IRAS 05405-0117. The
line can be seen in the same FFTS spectrum as the 372.42 GHz
H$_2$D$^+$ line.}
\label{figure:allspectra}
\end{center}
\end{figure*}

The Hanning smoothed spectra are shown in
Fig.~\ref{figure:allspectra}.  A summary of the observations with the
line parameters from Gaussian fits is presented in
Table~\ref{table:lineparameters}.

The H$_2$D$^+$ line is detected towards L183-S and $20^{\prime\prime}$
north and south of it.  The line in L183-S (0,0) with the best S/N is
single peaked, but suggests slight asymmetry. The FWHM
(0.42~kms$^{-1}$) is a little larger than expected from thermal
broadening at $T_{\rm kin} = 7$~K (0.28~km\,s$^{-1}$). Our L183
positions lie near positions included in the N-S oriented H$_2$D$^+$
strip observed at the Caltech Submillimeter Observatory (CSO) by
\cite{vastel2006}.  Their position $\Delta\delta=0^{\prime\prime}$
corresponds to the offset $(-4^{\prime\prime},+19^{\prime\prime})$
from L183-S. The antenna temperatures measured at APEX towards L183-S
$(0^{\prime\prime},0^{\prime\prime})$ and
$(0^{\prime\prime},+20^{\prime\prime})$, and the $3\,\sigma$ upper
limit $T_{\rm A}^* < 0.4$~K obtained towads L183-N (close to Vastel's
$\Delta\delta = +80^{\prime\prime}$) are consistent with the CSO data.
On the other hand, the line intensity at L183-S
$(0^{\prime\prime},-20^{\prime\prime})$, $T_{\rm A}^* \sim
0.9\pm0.2$~K, is surprisingly high in view of the fact that Vastel et
al. obtain antenna temperatures of $\approx 0.7$~K and 0.4~K towards
the offsets $(-4^{\prime\prime},-11^{\prime\prime})$ and
$(-4^{\prime\prime},-31^{\prime\prime})$ from L183-S, respectively.
This suggests that the H$_2$D$^+$ distribution peaks slightly south of
the dust emission peak but falls very steeply towards the southern end
of the dense ridge.
  
\begin{table}[ht]
\begin{center}
      \caption[]{Line parameters derived from Hanning smoothed spectra
with a velocity resolution of 98 ms$^{-1}$.}
         \label{table:lineparameters}
\begin{tabular}{lllll} \hline\hline
Line/          &   $T_{\rm A}^*$  &  $v_{\rm LSR}$ &   $\Delta v$(FWHM) & 
RMS \\
Position       &    (K)           &  (kms$^{-1}$)   &   (kms$^{-1}$) & (K) \\ 
\hline
{\bf H$_2$D$^+(1_{10}-1_{11})$} &  &  & & \\ 
IRAS 05405-0117 &  -  &    &   & 0.03   \\
OriB9 E         & 0.13 & $9.11 \pm0.06$ & $0.56\pm0.12$ & 0.04  \\
OriB9 N         & 0.11 & $9.20 \pm0.06$ & $0.42\pm0.15$ & 0.03   \\
L183-N          &   -   &     &   & 0.13  \\
L183-S $(0^{\prime\prime},0^{\prime\prime})$ & 0.68  &  $2.42\pm0.02$ &
   $0.42\pm0.04$ & 0.05 \\
L183-S $(0^{\prime\prime},-20^{\prime\prime})$ & 0.90 & $2.47\pm0.03$ & 
   $0.41\pm0.06$ & 0.15 \\
L183-S $(0^{\prime\prime},+20^{\prime\prime})$ & 0.42 & $2.41\pm0.06$ &  
   $0.47\pm0.13$ & 0.10 \\
RCrA NW         &  -  &   &  & 0.10 \\ \hline
{\bf N$_2$H$^+(4-3)$}  &             &             &    &    \\
IRAS 05405-0117 &  0.37  &  $9.22\pm0.01$  & $0.34\pm0.04$  & 0.04   
\\ \hline
\end{tabular}
\end{center}
\end{table}

In OriB9, weak H$_2$D$^+$ lines with $T_{\rm A}^* \approx
0.12$~K at $v_{\rm LSR} \sim 9.1-9.2 $ km\,s$^{-1}$ are detected
towards the two starless condensations, OriB9 E and N. The
LSR velocities agree with those of the previously observed NH$_3$ 
(9.2 km\,s$^{-1}$ at OriB9 E and N) and N$_2$H$^+$ 
(9.2 km\,s$^{-1}$ at the IRAS position) lines.  An upper limit of
0.10~K ($3\sigma$) is obtained towards IRAS 05405-0117.  This
spectrum has, however, another line with $T_{\rm A}^* \sim 0.4$~K at
an about 250 MHz higher frequency. The most probable identification is
N$_2$H$^+(4-3)$ at 372.67 GHz in the signal band (USB, $v_{\rm LSR} =
9.2$ km\,s$^{-1}$).  We estimate from the N$_2$H$^+(1-0)$ data of
\cite{caselli1994} that the N$_2$H$^+$ column density towards
IRAS 05405-0117 is $\sim 6-8 \, 10^{12}$~cm$^{-2}$. The
intensity of the supposed N$_2$H$^+(4-3)$ line is consistent with
this value. This line is not detected in other spectra.

Towards RCrA NW we obtained a $3\sigma$ upper limit of 0.3~K. The
observations were made at low elevations. Furthermore, the system
temperature was higher than expected from the observing conditions,
probably because of problems with the integration of the ASC correlator
into the system.

\section{Column densities and abundances}

The observed H$_2$D$^+$ line has quadrupole hyperfine structure due to
the spins of D and the two H nuclei. The splitting is, however, very
small ($\Delta \nu \sim 80$~kHz, \cite{jensen1991}) compared with the
Doppler width at 10 K ($\Delta \nu \sim 420$~kHz), and we treat the
line as if it had a single component. 
We estimate the {\sl ortho}-H$_2$D$^+$ column densities, $N(o-{\rm
H_2D^+})$, in the same manner as done in \cite{caselli2003}. A lower
limit for the excitation temperature, $T_{\rm ex}$, is obtained from
the observed $T_{\rm A}^*$ by assuming that the line is optically
thick.  The kinetic temperature of the gas, $T_{\rm kin}$, sets an
upper limit.

Towards L183-S (0,0), the reasonable $T_{\rm ex}$ range is $6.3\,{\rm
K} < T_{\rm ex} < 10\,{\rm K}$ (allowing $T_{\rm kin}$ be slightly
higher than the dust temperature).  Assuming that $T_{\rm ex} = T_{\rm
dust} = 7$~K, and that the source fills the main beam uniformly, we
arrive at the values $\tau_0 = 1.25$ and $N(o-{\rm H_2D^+}) =
2.7\,10^{13}$~cm$^{-2}$.  The main beam efficiency, $\eta_{\rm MB}$,
is assumed to be 0.7.  This column density is very close to that found
by \cite{caselli2003} towards the centre of L1544. The peak H$_2$
column density towards L183-S derived from SCUBA data is
$9.1\,10^{22}$~cm$^{-2}$ (Kirk \& Ward-Thompson, 2006, private
communication).  Using this value we obtain the fractional {\sl
ortho}-H$_2$D$^+$ abundance $X(o-{\rm H_2D^+}) = 3\, 10^{-10}$.  By
varying $T_{\rm ex}$ in the given range, $N(o-{\rm H_2D^+})$ and
$X(o-{\rm H_2D^+})$ decrease (higher $T_{\rm ex}$, smaller $\tau_0$) or
increase (lower $T_{\rm ex}$, larger $\tau_0$) by a factor of three. The line
shape does not support, however, the idea of large opacities, and we
think the values corresponding to $T_{\rm ex} = 7$~K are the most
likely.

OriB9 is little studied and dust continuum measurements are not
available. The $T_{\rm kin}$ derived from ammonia is 10~K, and the
minimum $T_{\rm ex}$ from the line intensity is 4~K. Using the
assumption $T_{\rm ex}=7$~K, which is midway between the two extremes, we
get $\tau_0 \sim 0.13$ and $N(o-{\rm H_2D^+}) \sim 3.0\,10^{12}$
cm$^{-2}$ towards N and E. The ammonia column density in these
positions is $\sim 10^{15}$~cm$^{-2}$. Assuming that the fractional
NH$_3$ abundance is $3\,10^{-8}$ (e.g. \cite{harju1993}) we get
$X(o-{\rm H_2D^+}) \sim 1\, 10^{-10}$.  The line profiles with rather
poor S/N ratios do not exclude large optical thicknesses. The column
density range implied by the possible $T_{\rm ex}$ values is $\sim
1.0\, 10^{12} - 5\, 10^{13}$ cm$^{-2}$, and the derived fractional abundance
has the corresponding uncertainty. 

According to the model of \cite{walmsley2004}, the characteristic
steady-state value of the {\sl o/p}-ratio of H$_2$D$^+$ is $\sim 0.3$
in the density range $n_{\rm H_2} \sim 10^5 - 10^6$~cm$^{-3}$
appropriate for the objects of this study.  Adopting this {\sl
o/p}-ratio the total H$_2$D$^+$ abundances in L183-S and 
OriB9 become $\sim 1.3 \,10^{-9}$ and $\sim 4 \,10^{-10}$.

\section{Discussion}

The performance of APEX and its equipment, and the atmospheric
transmission on Chajnantor are found to meet very high
standards. The telescope is therefore likely to become a very
important tool for studies of molecular cloud interiors and star
formation using H$_2$D$^+$, and other 'difficult' molecules.

The present observations towards a small sample of dense cores with
some diversity of physical characteristics suggest that either an
elevated temperature (as in R CrA NW) or the presence of an embedded
star (as towards IRAS 05405-0117), even if there is little
evidence for star-cloud interaction, decreases the changes to find
H$_2$D$^+$.  The kinetic temperature, velocity dispersion and the
fractional H$_2$D$^+$ abundance in OriB9 are similar to those
in the prestellar dark cloud cores L1544, 16293E, and L183,
where strong emission of this line has been detected previously. The
masses and the central densities of these nearby cores are
$2-3\,M_\odot$ and $\sim 10^6\,{\rm cm}^{-3}$, respectively
(\cite{vastel2006} and references therein).  The total mass of the
OriB9 core estimated from ammonia is of the order of $100 \,
M_\odot$ (\cite{harju1993}), and the subcondensations seen in the
N$_2$H$^+$ map (\cite{caselli1994}) are likely to be an order of
magnitude more massive than L1544, 16293E, or L183.  A
(sub)millimetre continuum map is needed to confirm this. Nevertheless,
OriB9 seems to be capable of forming a massive star or a
dense cluster of low- to intermediate-mass stars, and is therefore
exceptional among sources detected in H$_2$D$^+$ so far.

This detection confirms the existence of very cold, quiescent, dense
cores in massive star forming regions. The previous evidence for such
objects is scarce.  Some infrared dark clouds (IRDCs) have gas
temperatures approaching 10~K and ammonia linewidths slightly below 1
km\,s$^{-1}$ (\cite{pillai2006}). Yet another massive core with these
characteristics has been recently found in the region of ISOSS
J18364-0221 (\cite{birkmann2006}).  The linewidths in these objects
are, however, clearly larger than in OriB9. On the other
hand, only a small fraction of IRDCs have been observed in spectral
lines to date.  Because compression leads to an intensified cooling
by molecules and dust, the collapse of all dense cores should be
preceded by a cold, quiescent phase. Very cold GMC cores may have
remained indiscernible because of their short life-time or the fact
that large-scale surveys are usually biased towards the presence of a
certain molecular species which might be depleted in the coldest
regions.

As discussed recently by \cite{flower2006}, the H$_2$D$^+$ abundance
depends inversely on the {\sl ortho}:{\sl para} ratio of H$_2$, which
is largest at early stages of core evolution. Consequently, a high
degree of deuterium fractionation is a sign of matured chemistry
characterized by a low {\sl o}/{\sl p} H$_2$ and probably a high
degree of molecular depletion. The {\sl ortho}-H$_2$D$^+$ detection
towards OriB9 suggests an evolved chemical stage and tells of
a longlasting prestellar phase. However, estimates of the {\sl o}/{\sl
p} ratio of H$_2$ and the degree of deuterium fractionation using
other tracers are needed to confirm this. Despite these obstacles, it
seems viable to use the 372 GHz H$_2$D$^+$ line together with
chemistry models to estimate timescales related to the early evolution
of massive cores.

\vspace{3mm}

We thank Dr. J.R. Pardo for making the ATM program available, and
Malcolm Walmsley and the anonymous referee for helpful comments on the
manuscript.  The Helsinki group acknowledges support from the Academy
of Finland through grants 1201269, 1210518 and 206049.

\end{document}